\documentclass{moriond}

\def\be{\begin{equation}}
\def\ee{\end{equation}}
\def\bea{\begin{eqnarray}}
\def\eea{\end{eqnarray}}

\newcommand{\Photo}{}
\begin{document}
\vspace*{4cm}
\title{Greybody factors as robust gravitational observables: insights into post-merger signals and echoes from ultracompact objects}

\author{Romeo Felice Rosato}

\address{Dipartimento di Fisica, Sapienza Università di Roma \& INFN, Sezione di Roma, Piazzale Aldo Moro 5, 00185, Roma, Italy}
\maketitle\abstracts{The quasinormal mode spectrum plays a central role in modeling the post-merger ringdown phase of binary coalescences of compact objects. However, its interpretation is subject to certain ambiguities. Motivated by a recently discovered connection between greybody factors and post-merger black hole signals, we investigate the robustness of greybody factors as gravitational observables, offering a complementary perspective to quasinormal mode analysis. We show that greybody factors are stable under small perturbations and are not affected by the specific ambiguities that limit the reliability of quasinormal modes. Furthermore, we demonstrate that greybody factors are equally relevant for characterizing the signals emitted by wormholes and other horizonless ultracompact objects, providing a natural explanation for the echoes observed in the time-domain response of such objects.}

\section{Introduction}
Binary black hole~(BH) mergers are ideal laboratories to explore the dynamics of Einstein's theory~\cite{Berti:2015itd,Berti:2018vdi,Cardoso:2019rvt}. While their theoretical relevance has been long recognized, the advent of gravitational-wave~(GW) detections~\cite{LIGOScientific:2016aoc} has made them directly accessible, allowing us to probe the most extreme regions of spacetime. A binary BH coalescence evolves through three main phases: inspiral, merger, and ringdown. During the inspiral, the BHs lose energy via GW emission and eventually merge into a single, new BH. The ringdown phase, describing the relaxation of this remnant, is typically modeled using quasinormal modes~(QNMs), which correspond to exponentially damped oscillations. Although QNMs underpin BH spectroscopy, they are known to be highly sensitive to small perturbations of the system~\cite{Nollert:1996rf,Jaramillo:2020tuu,Cheung:2021bol}. This sensitivity can obscure physical interpretation, although the early-time time-domain signal remains relatively robust. To address these limitations, greybody factors (GFs) have been proposed as complementary observables~\cite{Rosato:2024arw,Oshita:2023cjz}, and extended to exotic compact objects~\cite{Rosato:2025byu}. GFs encode spectral features of GW emission and are less sensitive to small perturbations, making them promising probes of both BHs and horizonless compact objects. We will focus on a Schwarzschild geometry, in $G=c=1$ units.

\section{Greybody factors in black hole ringdown}\label{sec:GFs}
At linear level, the Schwarzschild BH response to external perturbations can be described via the radial equation~\cite{Zerilli:1970se}:
\begin{equation}\label{Regge-Zerilli_source}
 \left[{d^2 \over dr_*^2} + \omega^2 - V_l(r)\right] X_{lm\omega}=S_{lm\omega}(r)\,,
\end{equation}
where $r_*$ is the tortoise coordinate, $\omega$ the frequency, and $(l,m)$ are spherical-harmonic indices. The potential $V_l$ and source $S_{lm\omega}$ differ between axial and polar modes~\cite{Zerilli:1970se}.

\begin{figure}
\centering
\includegraphics[width=0.8\linewidth]{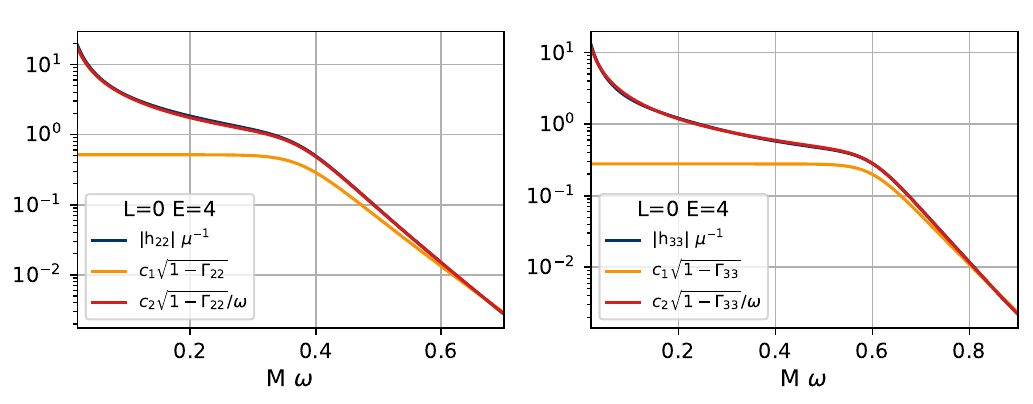}
\caption{Spectral amplitudes for $l=2,3$ modes from a point particle with $E=0$ and $L=4$~\protect\cite{Zerilli:1970se}. Left: $|h_{22}|$. Right: $|h_{33}|$. A model $\propto \sqrt{1-\Gamma_{lm}}/\omega^p$ is also shown for various $p$.}
\label{fig:h22} 
\end{figure}

We first consider the homogeneous equation. The greybody factor $\Gamma_{lm}(\omega)$ is the transmission coefficient of a scattering problem defined by:
\begin{equation}\label{boundary_refl/trasm}
\lim_{r_* \to -\infty}X_{lm\omega} \sim e^{- i \omega r_*}\,,\quad \lim_{r_* \to \infty}X_{lm\omega} \to A^{\rm in}_{lm\omega} e^{- i\omega r_*} + A^{\rm out}_{lm\omega} e^{+ i \omega r_*}\,.
\end{equation}
QNMs correspond to $A^{\rm in}_{lm\omega} = 0$. The reflectivity and transmissivity are defined as:
\begin{equation}
  \mathcal{R}_{lm}(\omega)=\left|\frac{A^{\rm out}_{lm\omega}}{A^{\rm in}_{lm\omega}}\right|^2\,, \quad \Gamma_{lm}(\omega)=\left|\frac{1}{A^{\rm in}_{lm\omega}}\right|^2\,, \label{RandT}
\end{equation}
with $\mathcal{R}_{lm} + \Gamma_{lm} = 1$. These quantities describe absorption and emission of the BH. For a Schwarzschld BH, $\Gamma_{lm}$ is a function of the frequency only depending on the mass of the compact object. References \cite{Oshita:2023cjz,Rosato:2024arw} show that, if the perturbation is induced by a plunging particle, the waveform amplitude $|h_{lm}(\omega)| \propto |X_{lm\omega}|$ is well approximated\,\footnote{Where $p$ takes different values depending on the source characteristics.} by $\sqrt{1 - \Gamma_{lm}(\omega)}/\omega^p$. Fig.~\ref{fig:h22} shows spectral amplitudes from a point particle plunging inside a BH, computed using Eq.~\ref{Regge-Zerilli_source} with source terms depending on initial energy $E$ and angular momentum $L$ of the particle~\cite{Zerilli:1970se}.
These results suggest that GFs could be directly observable in the GW spectrum and, being no-hair quantities\,\footnote{For a Kerr black hole, they only depend on mass and spin.}, may provide a robust tool for probing BH properties.

\section{Environmental effects}\label{Sec:enveffects}
QNMs are known to be spectrally unstable under small perturbations. In contrast, GFs remain stable. To explore this, we introduce a small P\"oschl--Teller-like bump in the potential~\cite{Cheung:2021bol} \footnote{For concreteness, we consider axial gravitational perturbations, but the analysis holds more generally.}:
\begin{equation}\label{eq:bump}
   V^\epsilon_l = \left(1 - \frac{2M}{r}\right)\left(\frac{l(l+1)}{r^2} - \frac{6M}{r^3}\right) + \frac{\epsilon}{M^2} {\rm sech}^2\left[\frac{r_* - c}{M}\right],
\end{equation}
with $\epsilon \ll 1$ and $c$ controlling amplitude and location. For $c \gg M$, the bump models environmental effects; for $c < 0$ and $|c| \gg M$, it mimics near-horizon structures due to quantum effects or matter overdensities.

\begin{figure*}[ht]
    \centering    \includegraphics[width=0.8\textwidth]{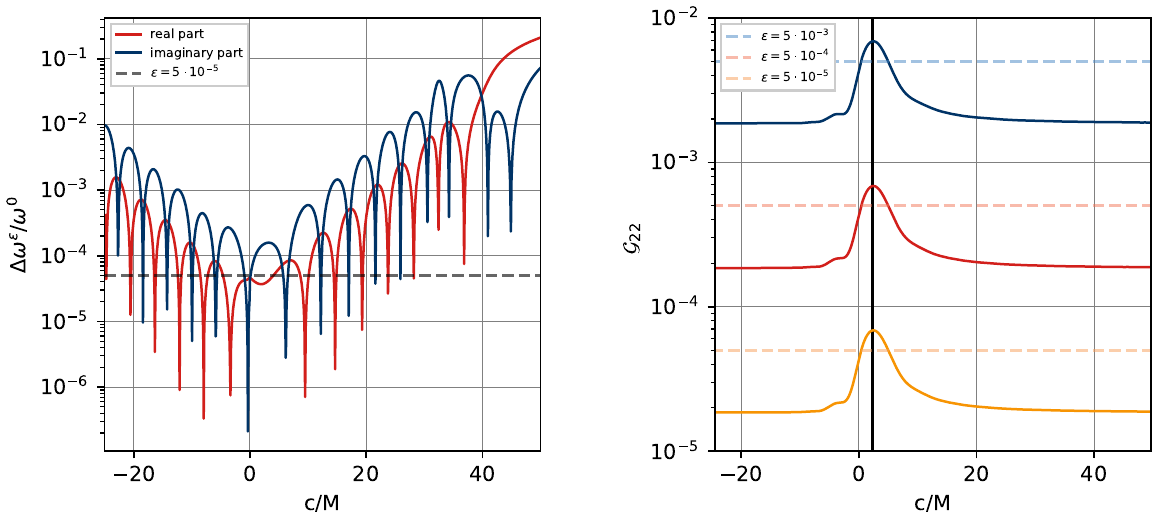}
    \caption{Instability of QNMs (left) vs. stability of GFs (right) under perturbations at $r_* = c$ with amplitude $\epsilon \ll 1$. Left: fixed $\epsilon = 5\times10^{-5}$. Right: varying $\epsilon$. The GF deviation peaks near the potential barrier ($c \approx 2.39 M$).}
    \label{fig:SchwPerturb}
\end{figure*}

Fig.~\ref{fig:SchwPerturb} compares the response of QNMs and GFs to the previous modification of the potential. The QNM deviation is quantified by
\begin{equation}
\frac{\Delta\omega^\epsilon_{R,I}}{\omega^0_{R,I}} = \left|\frac{\omega^\epsilon_{R,I}(c)}{\omega^0_{R,I}} - 1\right|,
\end{equation}
where $\omega^\epsilon_{R,I}(c)$ and $\omega^0_{R,I}$ are the perturbed and unperturbed real and imaginary parts, respectively.

For the GFs, we define the integrated deviation:
\begin{equation}\label{Deltal}
\mathcal{G}_{lm} = \frac{ \int_0^{\infty} |\Gamma^\epsilon_{lm}(\omega, c) - \Gamma_{lm}(\omega)| d\omega }{ \int_0^{\infty} \Gamma_{lm}(\omega) d\omega }\,.
\end{equation}

As seen in Fig.~\ref{fig:SchwPerturb}, QNMs show exponential sensitivity to $|c|$, leading to $\Delta\omega^\epsilon / \omega^0 \gg \epsilon$, confirming spectral instability~\cite{Jaramillo:2020tuu,Cheung:2021bol}. In contrast, GFs remain bounded, with $\mathcal{G}_{lm} < \epsilon$ as $|c| \to \infty$.

These findings confirm that greybody factors are robust against small perturbations, establishing them as stable and reliable observables for gravitational-wave phenomenology.
\section{Exotic compact objects}
The results of Sec.~\ref{sec:GFs} and~\ref{Sec:enveffects} have been extended to various ultracompact, non-rotating objects~\cite{Rosato:2025byu}. In this section, we focus on the wormhole case, modeled by two Regge--Wheeler potentials glued at the origin. The effective potential is given by $V_l(r_*) = \theta(r_*) W_l(r_* + r_*^0) + \theta(-r_*) W_l(-r_* + r_*^0)$, where $W_l(r_*)$ denotes the Regge--Wheeler potential~\cite{Bueno_2018}.

\begin{figure*}
\begin{minipage}{0.5\linewidth}
\centerline{\includegraphics[width=0.8\linewidth]{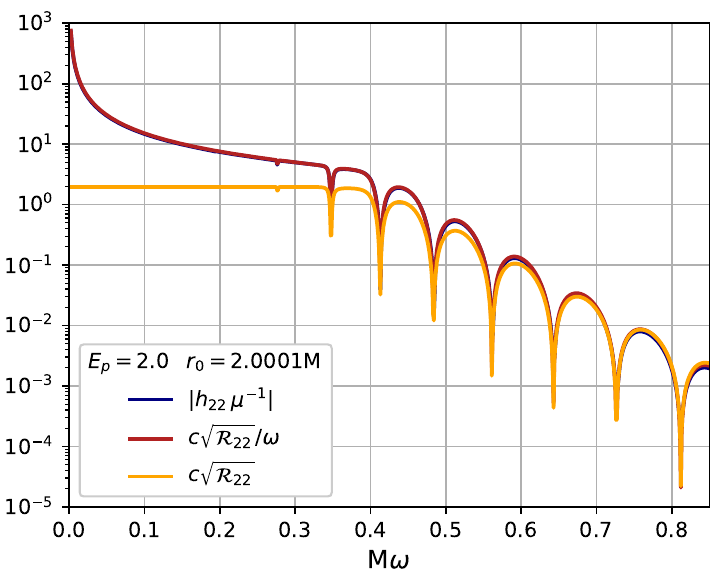}}
\end{minipage}%
\begin{minipage}{0.5\linewidth}
\centerline{\vspace{-0.3cm}\includegraphics[width=0.8\linewidth]{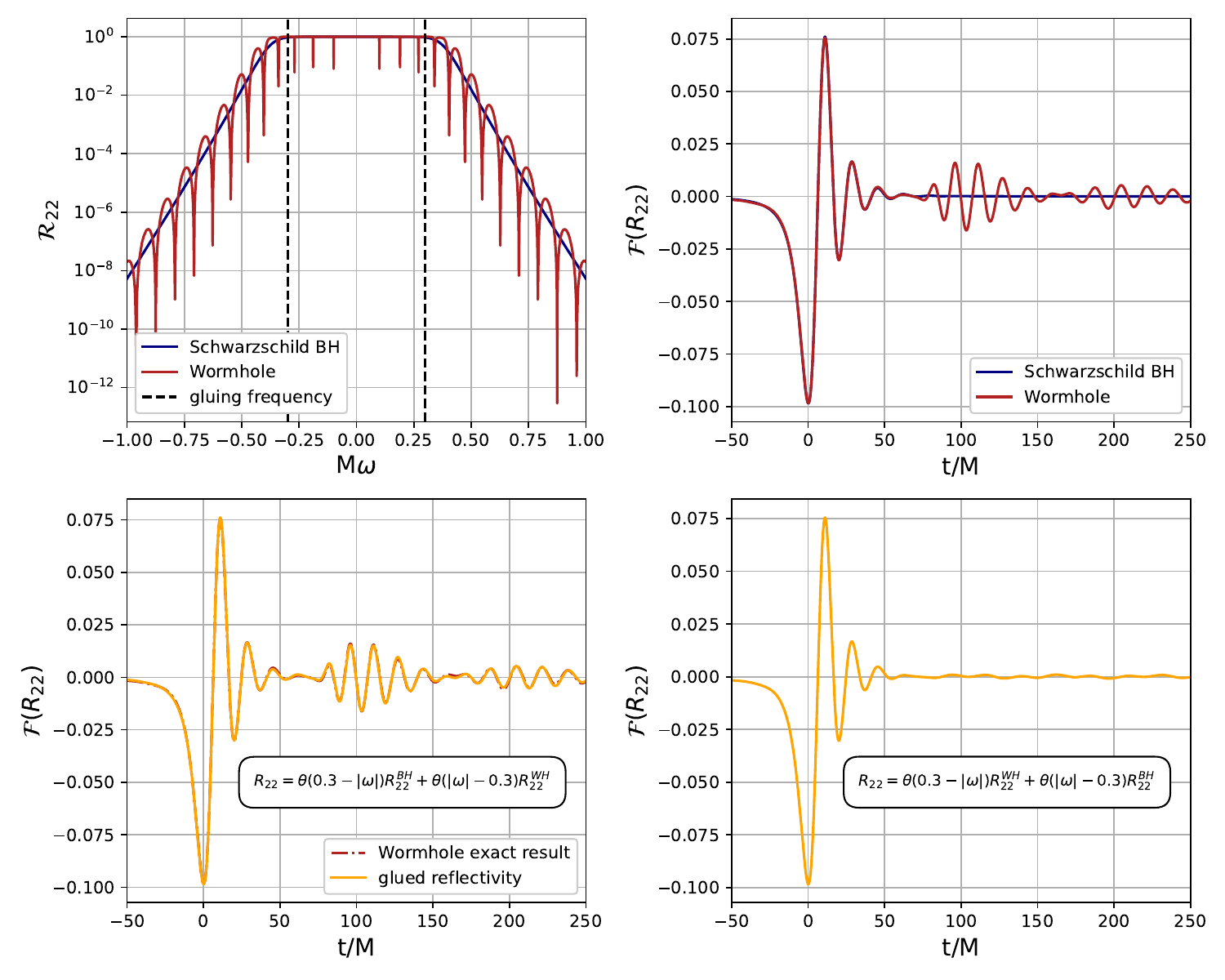}}
\end{minipage}
\caption[]{Left: spectral amplitude for $l=2$ emitted by a point particle with mass $\mu$ in radial infall, with specific energy $E_p$. For comparison, two models $\propto \sqrt{\mathcal{R}_{lm}}/\omega$ and $\propto \sqrt{\mathcal{R}_{lm}}$ are shown. The model $\propto \sqrt{\mathcal{R}_{lm}}/\omega$ accurately captures the spectral amplitude across all frequencies. Right: Fourier transform of the reflectivity for a Schwarzschild-like wormhole with gluing radius $r_0 = 2.0001M$ (red), compared to a Schwarzschild BH (blue). Echoes are clearly visible in the wormhole case.}
\label{fig:Wormhole}
\end{figure*}

The scattering problem is treated with the same boundary conditions used for the BH case. Due to the symmetric potential structure, a cavity forms at the center, producing resonances and oscillations in the final reflectivity, as shown in the left panel of Fig.~\ref{fig:Wormhole}.

While we omit technical details here, Ref.~\cite{Rosato:2025byu} demonstrates that:
\begin{itemize}
    \item Again, the spectral amplitude of the emitted gravitational waves $|h_{lm}(\omega)|$ is well approximated by $\sqrt{1 - \Gamma_{lm}(\omega)}/\omega^p$, as showed in Fig.~\ref{fig:Wormhole} left panel.
    \item At high frequencies, the system supports reflectionless scattering modes, corresponding to waves that are fully transmitted. This behavior can be derived analytically.
    \item Oscillations in the high-frequency reflectivity lead to gravitational-wave echoes in the time-domain response of wormholes. An example is provided in Fig.~\ref{fig:Wormhole} (right panel), comparing the Fourier transform of the wormhole reflectivity to that of a Schwarzschild BH.
\end{itemize}

\section{Conclusions}
GFs appear to be promising candidates as gravitational-wave observables, offering insights into the nature of both black holes and exotic compact objects. Future developments may enable the construction of novel tests of gravity based on GFs, providing a complementary approach to QNMs.
\section*{Acknowledgments}
I would like to thank my PhD advisor, Paolo Pani, as well as all of my collaborators: K. Destounis, S. Biswas, and S. Chakraborty.
\section*{References}
\bibliography{moriond}


\end{document}